\documentclass[sigconf]{acmart}

\usepackage{subfigure}
\usepackage{algorithm}
\usepackage[noend]{algpseudocode}
\usepackage{multirow}
\usepackage{tcolorbox}

\usepackage{tabu}
\usepackage{multirow}

\usepackage{listings}
\usepackage{multirow}

\makeatletter

\let\comment\@undefined
\let\endcomment\@undefined
\makeatother

\usepackage[commandnameprefix=ifneeded, final]{changes}
\usepackage[table]{xcolor}
\AtBeginDocument{%
  }

\begin{document}

\title{Backdoor Attacks on Fault Detection and Localization in Cyber-Physical Systems}


\author{Abile Jean, Kuniyilh S}

\settopmatter{printacmref=false}
\setcopyright{none}
\renewcommand\footnotetextcopyrightpermission[1]{}

\begin{abstract}
Cyber-Physical Systems (CPS) integrate sensing, communication, computation, and control to support critical infrastructure, including smart grids, industrial automation, and control systems. In the electrical utility domain, various controllers are used in CPS to ensure the system detects and recovers from faults, such as voltage fluctuations, and to perform load balancing in distribution systems. Machine learning- and deep learning-based fault detection and localization frameworks have recently gained significant attention in CPS for their ability to identify anomalies and operational failures in real time. However, these intelligent models are vulnerable to adversarial machine learning attacks, particularly backdoor attacks. In a backdoor attack, an adversary injects malicious patterns into the training data so that the model behaves normally most of the time but produces attacker-controlled outputs when triggered by specific patterns. This paper investigates the threat of backdoor attacks against fault detection and localization mechanisms in recent ML pipelines used in modern CPS systems. We define these threats and explore how they can be realized by designing triggers and evaluating their success in the CPS domain. Our experiments show the attack is successful even with 10\% of poisoning.
\end{abstract}



\maketitle
\section{Introduction}

Cyber-Physical Systems (CPS) have become foundational components of modern critical infrastructure, integrating computational intelligence to enable automation, monitoring, and control across industrial systems, transportation networks, smart grids, healthcare devices, and manufacturing environments~\cite{lee2015cyber}. The increasing adoption of machine learning (ML) and deep learning (DL) in CPS has significantly improved fault detection and localization capabilities by enabling predictive maintenance, anomaly detection, and adaptive operational control~\cite{cardenas2008ics_security, melnyk2025hardware, cps1}.

Fault detection and localization are essential functionalities in CPS~\cite{survey2024process, chauhan2023high, survey2026fdd, jin2022novel}. Fault detection identifies abnormal operational conditions, while fault localization determines the origin and location of faults within the system. Intelligent fault diagnosis systems are extensively employed in smart manufacturing, power systems, autonomous vehicles, and industrial Internet of Things (IIoT) infrastructures~\cite{fault1, thomas2023cnn, najafzadeh2024fault}.

Despite these advantages, ML-enabled CPS are vulnerable to adversarial attacks~\cite{chathoth2025pcap, flcps2025, koyatanpcap}. Among these threats, backdoor attacks are particularly dangerous because they remain stealthy during normal operation. In a backdoor attack, an attacker injects poisoned data samples containing specific triggers into the training dataset. The trained model performs correctly on benign inputs but misclassifies inputs containing the trigger.
In CPS, successful backdoor attacks can suppress alarms, misidentify fault locations, or trigger false fault responses. Such attacks can lead to catastrophic physical consequences, including equipment damage, process instability, financial loss, and risks to human safety.

This paper presents a novel fault detection and localization technique and analyzes backdoor attacks targeting fault detection and localization systems in CPS. The major contributions of this paper are summarized as follows:

\begin{itemize}
    \item We propose an LLM-based contrastive CPS fault detection and localization architecture.
    \item We design and implement a novel backdoor trigger generator for an LLM-based fault detection and localization pipeline for CPS.
    \item We evaluate the backdoor threat and injection strategies under various conditions and measure the sensitivity to various design parameters.
\end{itemize}

\section{Background}

\subsection{Cyber-Physical Systems}

\begin{figure}
    \centering
    \includegraphics[width=0.8\linewidth]{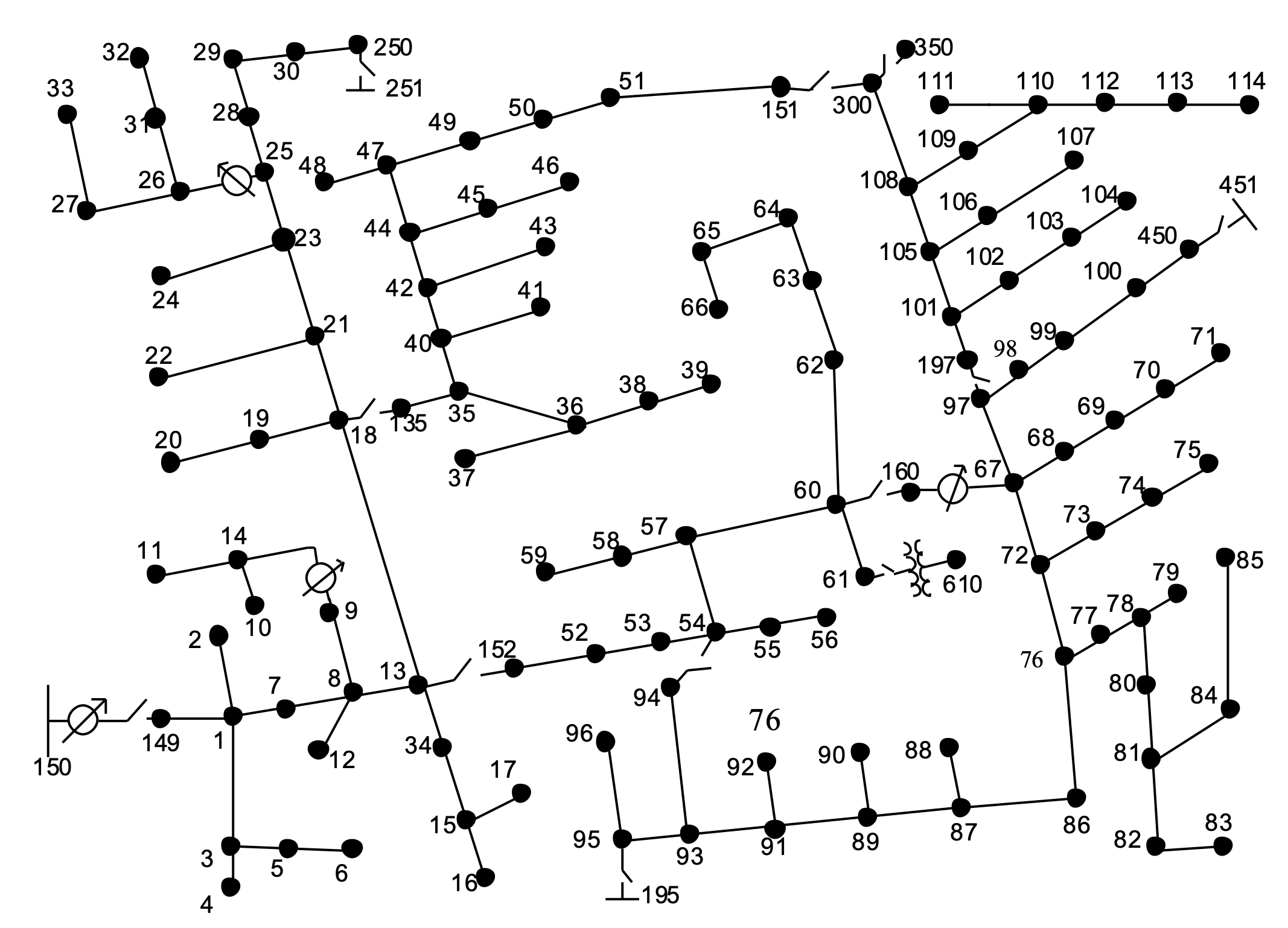}
    \caption{IEEE 123-bus system diagram}
    \label{fig:IEEE123}
\end{figure}

Cyber-Physical Systems integrate computational resources with physical processes through sensing, communication, and actuation. CPS are deployed in various applications such as 
Smart grids, Industrial control systems, Autonomous vehicles, Smart healthcare systems, Intelligent transportation systems.
CPS systems are typically deployed in critical infrastructure such as substations and nuclear plants, which makes researchers rely on simulated models to obtain real data for research and experiments~\cite{cps1}.
IEEE bus systems are used by researchers to implement new ideas and concepts. CPS data derived from the IEEE bus system is commonly used for research and evaluation purposes~\cite{elgindy2022sample}. The IEEE 123-bus test feeder is a standard radial distribution network model used by power engineers to test and validate simulations, volt-var optimization (VVO), and distribution system state estimation algorithms~\cite{soofi2023unleashing}. It represents an active \(4.16 \text{ kV}\) unbalanced system with multiple voltage regulators, shunt capacitors, and overhead/underground lines as shown in Figure~\ref{fig:IEEE123}. Because of its unbalanced loading conditions and voltage constraints, the 123-bus system is a frequent benchmark in academic and industrial power research. Primary use cases include simulating the integration of distributed energy resources (DER).
Anomaly and intrusion detection are instrumental components of model-connected devices and sensors deployed in CPS and IoT networks, as they help safeguard these systems~\cite{chathoth2021federated, liu2021anomaly, chathoth2022differentially}.

\subsection{Volt-Var attack}

Volt–Var control (VVC) is a critical component in distributed energy systems that automatically adjusts reactive power output based on measured voltage levels to keep them within nominal limits (e.g., 0.95–1.05 p.u.). This helps maintain voltage stability, reduce losses, and ensure efficient power delivery~\cite{selim2025robust}.
A Volt–Var attack is a type of cyberattack on a VVC system, where an attacker intends to increase or decrease the the distribution grids voltage beyond its normal range or outside its safe range by adversely controlling the or adjusting reactive power from devices like capacitor banks, roof top PV cells, on-load tap changers (OLTCs), electric vehicle chargers, and other distributed energy resources (DERs)~\cite{fragkos2025graphllm}.

\subsection{Fault Detection and Localization}

Fault detection refers to identifying abnormal operating conditions, while fault localization determines the source of the fault, such as a volt-var attack on a controller in CPS~\cite{najafzadeh2024fault}. AI-driven approaches employ supervised and unsupervised learning, as well as deep neural networks, for classification and diagnosis.
A typical fault diagnosis model can be represented as:

\[
y = f(x; \theta)
\]

where:
\begin{itemize}
    \item $x$ represents sensor observations,
    \item $\theta$ denotes model parameters,
    \item $y$ denotes predicted fault labels.
\end{itemize}

In the event of a Volt-VAR-Attack, the goal is to detect and localize nodes exhibiting anomalous or fault behavior. 



\section{ML-Based Fault Detection and Localization in CPS}

Recent advances in machine learning (ML) and deep learning (DL) have significantly transformed fault detection and localization mechanisms in Cyber Physical Systems (CPS)~\cite{fault1}. Modern CPS environments generate large-scale heterogeneous sensor data streams that traditional model-based methods often struggle to process effectively. Consequently, data-driven ML approaches have become dominant due to their scalability, adaptability, and capability to learn complex nonlinear system behaviors.
Recent survey literature highlights the rapid evolution of intelligent fault diagnosis frameworks for industrial CPS, smart grids, autonomous transportation systems, and Industrial Internet of Things (IIoT) infrastructures. \cite{survey2024process,survey2026fdd}

\subsection{Traditional ML-Based Fault Detection}

Earlier ML-based fault diagnosis systems primarily employed classical supervised learning methods such as Support Vector Machines (SVM), Random Forests (RF),k-Nearest Neighbors (kNN), Decision Trees, and Bayesian Networks~\cite{survey2024process}.
These techniques were widely used for equipment health monitoring, predictive maintenance, and anomaly detection in industrial CPS. Traditional ML methods generally require handcrafted feature extraction using signal processing techniques such as wavelets and Fourier transforms, as well as statistical feature engineering.
A comprehensive survey by Melo \textit{et al.} analyzed data-driven process monitoring and fault diagnosis approaches across industrial systems and highlighted the transition from statistical process control toward AI-enabled intelligent diagnostics~\cite{survey2024process}.

\subsection{Deep Learning for Fault Detection}

Deep learning has significantly improved fault detection accuracy by enabling automatic feature extraction from raw sensor streams. Convolutional Neural Networks (CNNs), Recurrent Neural Networks (RNNs), Long Short-Term Memory (LSTM), and Transformer architectures are increasingly adopted for CPS monitoring~\cite{thomas2023cnn}.
Deep architectures provide several advantages, such as Automatic feature learning, temporal pattern recognition, and multi-modal sensor fusion
CNNs are extensively used for vibration-based machinery fault diagnosis, while LSTM networks are highly effective for temporal anomaly detection in dynamic CPS environments.
Recent surveys indicate that hybrid CNN-LSTM frameworks outperform conventional methods for predictive maintenance and anomaly localization by jointly modeling spatial and temporal system behavior~\cite {survey2026fdd}.


While ML improves CPS reliability, recent literature also highlights growing vulnerabilities to adversarial, poisoning, and backdoor attacks.
Recent CPS security surveys emphasize that ML-enabled CPS must be designed with integrated resilience against adversarial machine learning attacks \cite{resilientml2021,evasion2024}.
Recent literature indicates a growing shift toward hybrid AI architectures that combine physics-informed learning, graph reasoning, explainability, and cybersecurity-aware model design for next-generation CPS fault management systems~\cite{survey2026fdd,anomaly2025, peng2025log}.
Building on prior work, we propose an LLM-based fault detection and localization system that leverages the flexibility of advanced computing and natural language processing.

\section{Backdoor attacks}


Backdoor attacks are poisoning-based adversarial attacks in which an attacker manipulates the training process by inserting maliciously crafted samples~\cite{gu2017badnets, gao2023backdoor, chathoth2025pcap, liang2024badclip, yang2022backdoorcl}. The objective is to implant hidden malicious behavior within the model.
The poisoned dataset $D'$ can be represented as:

\[
D' = D \cup D_p
\]

where:
\begin{itemize}
    \item $D$ is the original clean dataset,
    \item $D_p$ is the modified dataset containing trigger patterns.
\end{itemize}

The attacker introduces a trigger function on each data in the modified dataset $D_p$ :

\[
x_t = x + \delta
\]

where:
\begin{itemize}
    \item $x_t$ is the triggered input,
    \item $\delta$ is the trigger pattern.
\end{itemize}

The attacker's objective function $f()$ is designed to misclassify data with a trigger as the target class while maintaining normal performance on benign samples. 

\[
f(x_t) = y_{target}
\]

Backdoor attacks in CPS involve making changes that are feasible in real-world industrial control systems or the energy domain, where the laws of physics must be taken into account.

There are several types of backdoor attacks defined in the literature.

\textbf{False Negative Attacks:} The attacker suppresses the detection of real faults. The ML model incorrectly classifies a fault state as normal.

\[
f(x_t) = y_{normal}
\]

This attack is dangerous because it allows faults to propagate undetected.

\textbf{False Positive Attacks:} The attacker intentionally triggers false alarms by misclassifying benign as fault, as represented by

\[
f(x_t) = y_{fault}
\]

False positives can cause operational disruption and unnecessary diagnostics and maintenance efforts.

\textbf{Stealthy Persistent Attacks:} Stealthy attacks maintain high model accuracy on clean data while remaining dormant until the trigger is activated.
The optimization objective becomes:

\[
\min_{\theta} \; L_{clean} + \lambda L_{trigger}
\]

where:
\begin{itemize}
    \item $L_{clean}$ is the benign loss,
    \item $L_{trigger}$ is the trigger activation loss,
    \item $\lambda$ controls attack strength.
\end{itemize}

\textbf{Localization Misguidance:}
Fault localization systems identify faulty components using graph-based reasoning, sensor fusion, and spatial-temporal analysis.
Here, the attacker manipulates the localization output:

\[
f(x_t) = l_{target}
\]

where $l_{target}$ represents an incorrect fault location.

\textbf{Multi-Node Trigger Attacks:}
Distributed CPS environments allow attackers to activate triggers across multiple sensors simultaneously.
Impacts include incorrect fault isolation, cascading failures, and reduced situational awareness.

Our work introduces a novel stealthy clean-label attack in which the model learns to modify clean data so it is misclassified as the target class by introducing a trigger tailored to the attacker's goal.
We explain the design and the model details in the following sections.

\begin{figure}[t]
    \centering
    \includegraphics[width=1\linewidth]{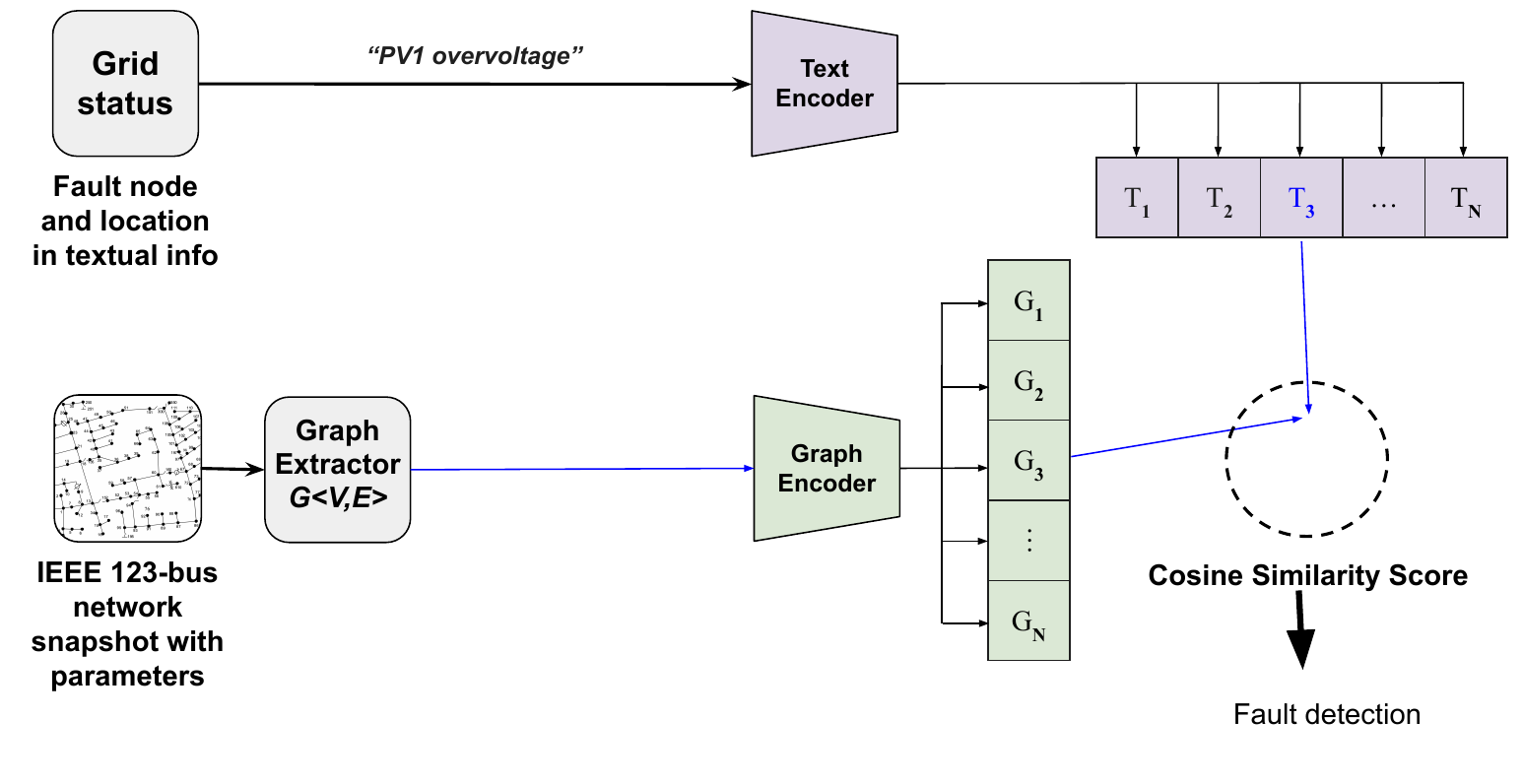}
    \caption{Fault detection model training pipeline}
    \label{fig:IEEE123-Fault-model}
\end{figure}

\begin{figure}[t]
    \centering
    \includegraphics[width=1\linewidth]{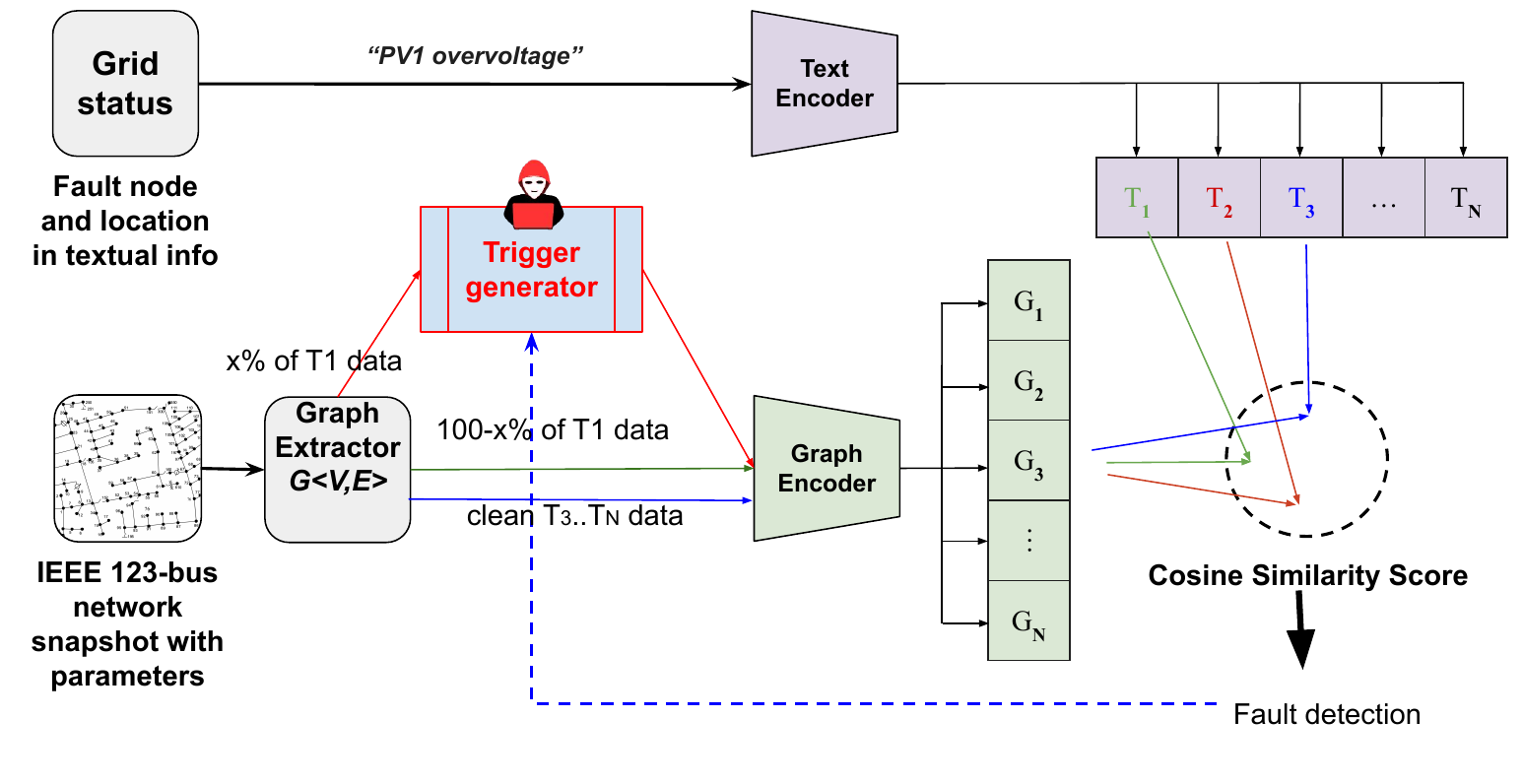}
    \caption{Fault detection backdoor model training pipeline}
    \label{fig:IEEE123-Fault-BD-training}
\end{figure}

\section{Backdoor attack on CPS fault localization}
In this section, we first discuss our threat model and then consider the IEEE 123-bus system-based CPS fault detection and localization model within the scope of this research. After this, we describe the backdoor training and attack workflow in detail.

\subsection{Threat Model}

We consider a threat model in which an adversary has access to training data or the supply chain and can manipulate node data streams or communication packets.
We also assume that the adversary has partial knowledge of ML architectures or the system, can be partially involved in the model training process, or can provide input data and receive loss or error information during training. We assume that some nodes in the CPS system are compromised or have vulnerabilities that allow them to process data maliciously before passing it to a classifier model. We also assume that such malicious nodes are involved in the training process, either directly or by sharing their data.
Attack goals include suppressing genuine fault alarms, triggering false fault localization, inducing unsafe operational decisions, and creating stealthy system instability.

\subsection{Design}

We propose an LLM-based fault detection model for an IEEE 123-bus Cyber-Physical System (CPS).  We leverage contrastive learning, particularly in a multimodal framework similar to CLIP~\cite{chathoth2025dynamic, radford2021learning, chathoth2025privclip, khosla2020supcon, chathoth2026contrastive}.In this approach, data is represented as a time-series graph, with vertices and edges denoting power parameters and flow directions. Additionally, we include a textual description of fault localization as the label for each snapshot of the time-series data~\cite{wu2025dual}.  Contrastive Graph Auto-Encoder (CGAE) and Contrastive Variational Graph Auto-Encoder (CVGAE) are recently introduced graph encoders for contrastive learning~\cite{mrabah2023contrastive}. We use CGAE as our graph encoder.

These multimodal data are used to train a model via a contrastive approach that employs both text and graph encoders, as illustrated in Figure~\ref{fig:IEEE123-Fault-model}. Once the model is trained, it can effectively identify the most similar pairs in a contrastive manner by projecting them into a latent space. The pair with the highest similarity score is considered the correct classification.
We chose this design because it offers greater flexibility compared to traditional deep learning models. This flexibility arises from the ability to use natural language descriptions for various system states. Furthermore, contrastive learning ensures that the encoders are projected into the latent space based on similarity, thereby reducing the number of input samples required for training and enabling few-shot learning.

We will now explain the backdoor technique using the previously described method for LLM-backed fault detection, in line with the identified threat model. This explanation will cover both the backdoor training and attack phases separately.

\textbf{Backdoor training phase:}
We introduce a backdoor generator or trigger generator in the training pipeline such that some portion of the input data is altered by the trigger generator, as shown in Figure~\ref{fig:IEEE123-Fault-BD-training}.
The trigger generator we use is an autoencoder, consisting of an encoder and a decoder designed to reconstruct the input data. Additionally, autoencoders can be employed to create backdoors, as they can learn the underlying data representation and reconstruct it without losing essential information~\cite{chathoth2024dynamic, mrabah2023contrastive}.
The trigger generator is trained based on the contrastive loss, as well as the generator divergence loss, as represented by the KL divergence loss for the autoencoder~\cite{van2014renyi, chathoth2024dynamic}. The goal of the trigger generator is to minimize the change while achieving the attacker's goal of classifying the target class, thereby enabling a stealthy attack.
During training, as assumed in our threat model, a small portion of the data corresponding to the target class is passed through the trigger generator along with the original textual label for model training. Recall that the target class is the attacker's preferred misclassification class when a backdoor is present. The goal here is to make the model predict the class label regardless of whether a trigger is added. The attacker also controls the trigger generator to apply minimal changes, guided by the loss functions, to make the trigger stealthy. For simplicity of demonstration, we consider only binary classification in most scenarios, where one class is fault, and the other is clean. During the contrastive learning phase, the latent representations of both the textual and graph encoders are projected into a latent space based on similarity scores, pulling dissimilar representations apart and maintaining a distinct space between the clean and fault projections, as highlighted by the green, blue, and red arrows.
By the end of the training process, as demonstrated in Figure~\ref{fig:IEEE123-Fault-BD-training}, the graph encoder, the trigger generator, and the contrastive model will be trained to perform their intended tasks. The loss functions are leveraged to get optimal performance. We use a pretrained graph extractor and text encoder in the pipeline.

\begin{figure}[t]
    \centering
    \includegraphics[width=1\linewidth]{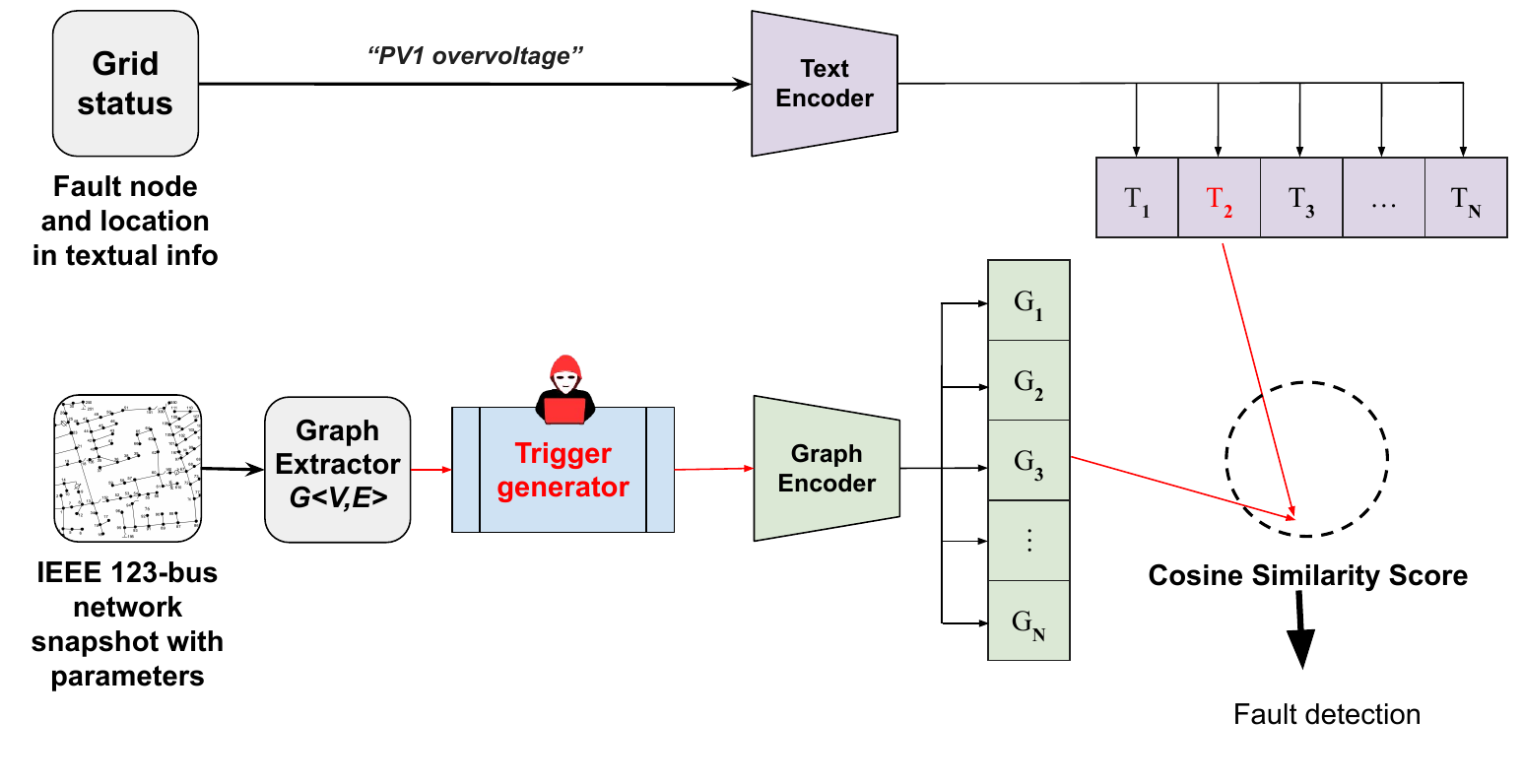}
    \caption{Fault detection backdoor attack on target data}
    \label{fig:IEEE123-Fault-attack}
\end{figure}

\begin{figure}[t]
    \centering
    \includegraphics[width=1\linewidth]{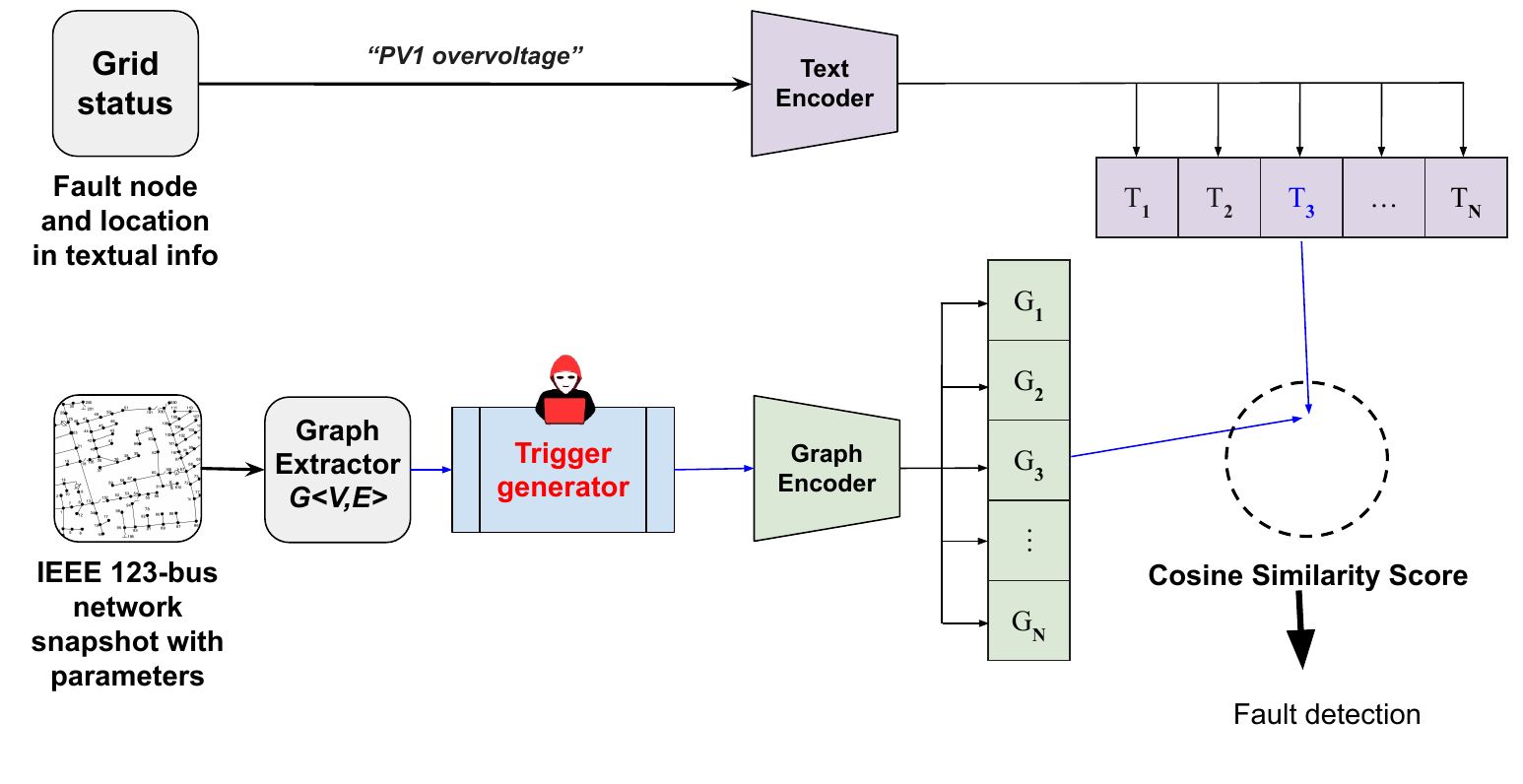}
    \caption{Fault detection backdoor attack on non-target data}
    \label{fig:IEEE123-Fault-clean}
\end{figure}

\textbf{Backdoor attack phase:}
Once the models in the pipeline are trained jointly, the model can be deployed at various nodes in the CPS. During the attack phase, an attacker who intends to misclassify the input to a target class or their intended class to be misclassified, for example, data corresponding to a fault class to be classified as a clean class, first activates the trigger by passing the data through the trained trigger generator, and the modified data with the dynamically generated trigger embedded is sent to the model for classification.
The attack workflow is captured in Figure ~\ref{fig:IEEE123-Fault-attack}. Since the model was trained to classify data with the presence of a trigger as the target class, the attack will be successful here.
Without a trigger, the model behaves normally as indicated in Figure~\ref{fig:IEEE123-Fault-clean}, where the clean data is sent to the classifier, and it classifies the model normally.

\section{Evaluation}
We evaluate the performance of our backdoor technique on simulated data.
We first explain the model architecture and the dataset used for our experiments.

\textbf{Model details:}
The classifier is trained to detect the fault and localization using a few-shot contrastive learning method similar to the CLIP model. Based on the similarity score, the label is assigned to the prediction.
We use the CGAE as our graph encoder~\cite{mrabah2023contrastive} and the pretrained text encoder from OpenAI's CLIP pipeline~\cite{radford2021learning}. 
The trigger generator is an autoencoder, and we use the same structure as used in ~\cite{chathoth2024dynamic}.

\textbf{Dataset:}
In order to generate the training and test datasets, faults are
simulated for all buses in the system.
Two types of faults are considered. One is overvoltage and voltage drops. Additionally, a normal operating dataset where the VVC adjusts the voltage within the safe range of the IEEE 123-bus system.
We modified the data from the Open Distribution System Simulator(OpenDSS) IEEE 123-bus system for our simulation~\cite{hariri2017open}. 
The OpenDSS is a comprehensive electrical power system simulation tool primarily for electric utility power distribution systems. 
We split the training and testing datasets into various combinations (90-10, 80-20, 70-30) to evaluate the few-shot capability.
Since we use contrastive learning, we also expect the model to learn to classify accurately even with a small portion of training data.

In the following subsections, we provide the experimental results evaluation.
\subsection{Impact on utility}
To study the impact of the backdoor generator in the model training pipeline, we compare the performance of the clean and backdoored models. In our experiments, we select one class at a time as the target, run tests at three target misclassification rates (10\%, 20\%, and 30\%), and compute the average performance score across accuracy, precision, recall, and F1-score.
We present the model comparison results in Table~\ref{tab:utility}.
As shown, the backdoor performance slightly dropped in the backdoored model because the trigger generator is involved, causing the model to lose some utility at the expense of trigger injection. However, overall, the loss of utility is minimal and can be improved by finetuning the model training process or data selection.

\begin{table}
    \caption{Performance comparison between clean model and backdoor model on clean data}
    \centering
    \begin{tabular}{|c|c|c|c|c|}\hline
        Model & Accuracy & Precision &  Recall & F1 \\\hline
         Clean model& 0.87 &  0.88& 0.88 & 0.88\\\hline
         Backdoor model& 0.84 & 0.84 & 0.84 & 0.84\\ \hline
    \end{tabular}

    \label{tab:utility}
\end{table}



\subsection{Backdoor performance sensitivity}

\begin{figure}[t]
    \centering
    \includegraphics[width=0.9\linewidth]{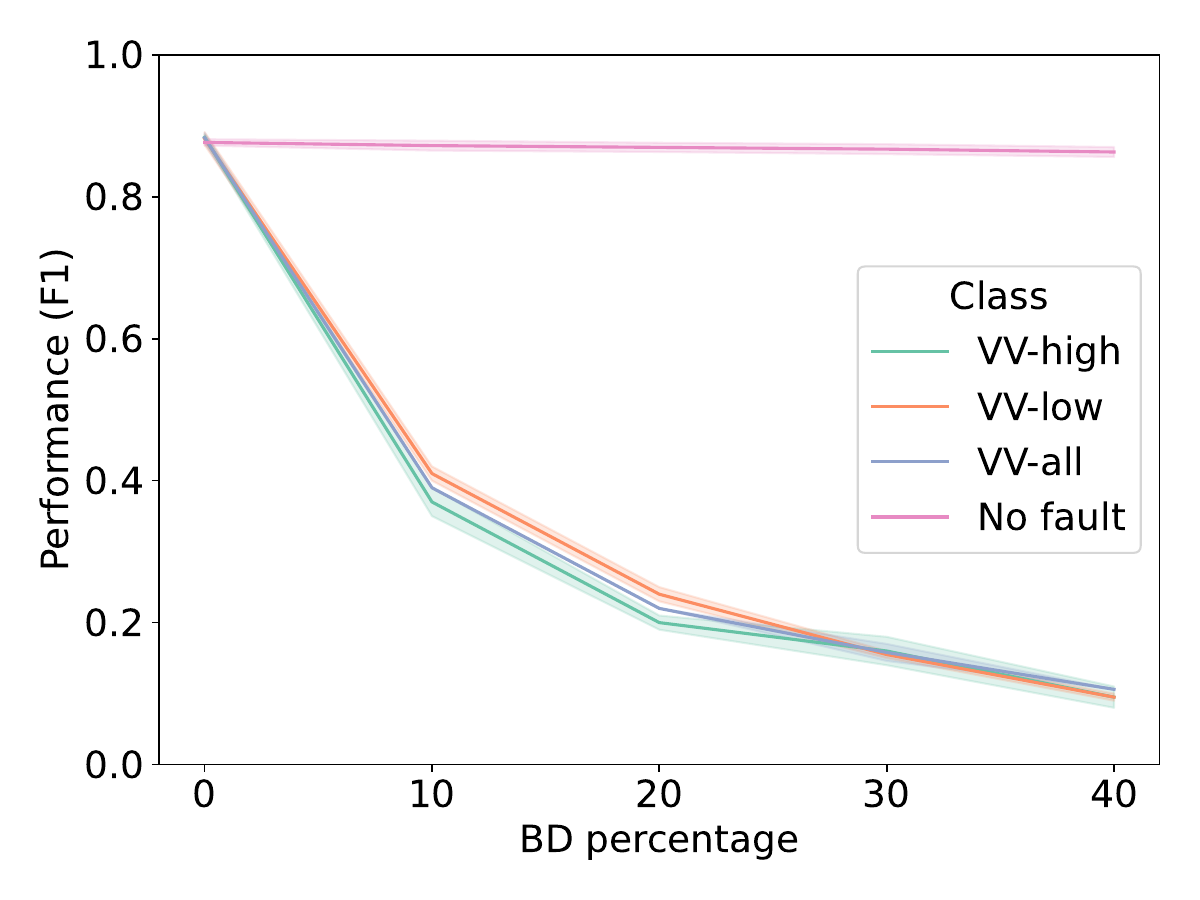}
    \caption{Backdoor performance sensitivity to BD percentage changes on each class}
    \label{fig:BackdoorPerf}
\end{figure}

Next, we discuss the sensitivity of backdoor performance on the backdoor percentage. We define the backdoor percentage as the proportion of the backdoor sample relative to the total training sample used in model training. We plot the result in Figure~\ref{fig:BackdoorPerf}, showing that clean data performance is consistent and not affected by the backdoor, since we consider clean data as the target class for the attacker, meaning that backdoor data samples taken from the No-fault class are used for the trigger generator and model training. Once the model is trained, any other class data, when embedded with a trigger, is classified as the No-fault class, which is the target class assumed in this experiment.
We further notice that as the backdoor percentage increases, the attack becomes more successful, as shown by the low F1-Scores for each fault class while keeping the No-fault class's F1-score consistent, thereby keeping the model stealthy.
This experiment demonstrates that the backdoor generator can be improved by supplying more training samples.




\subsection{Performance comparison}







\begin{figure}[t]
    \centering
    \includegraphics[width=0.9\linewidth]{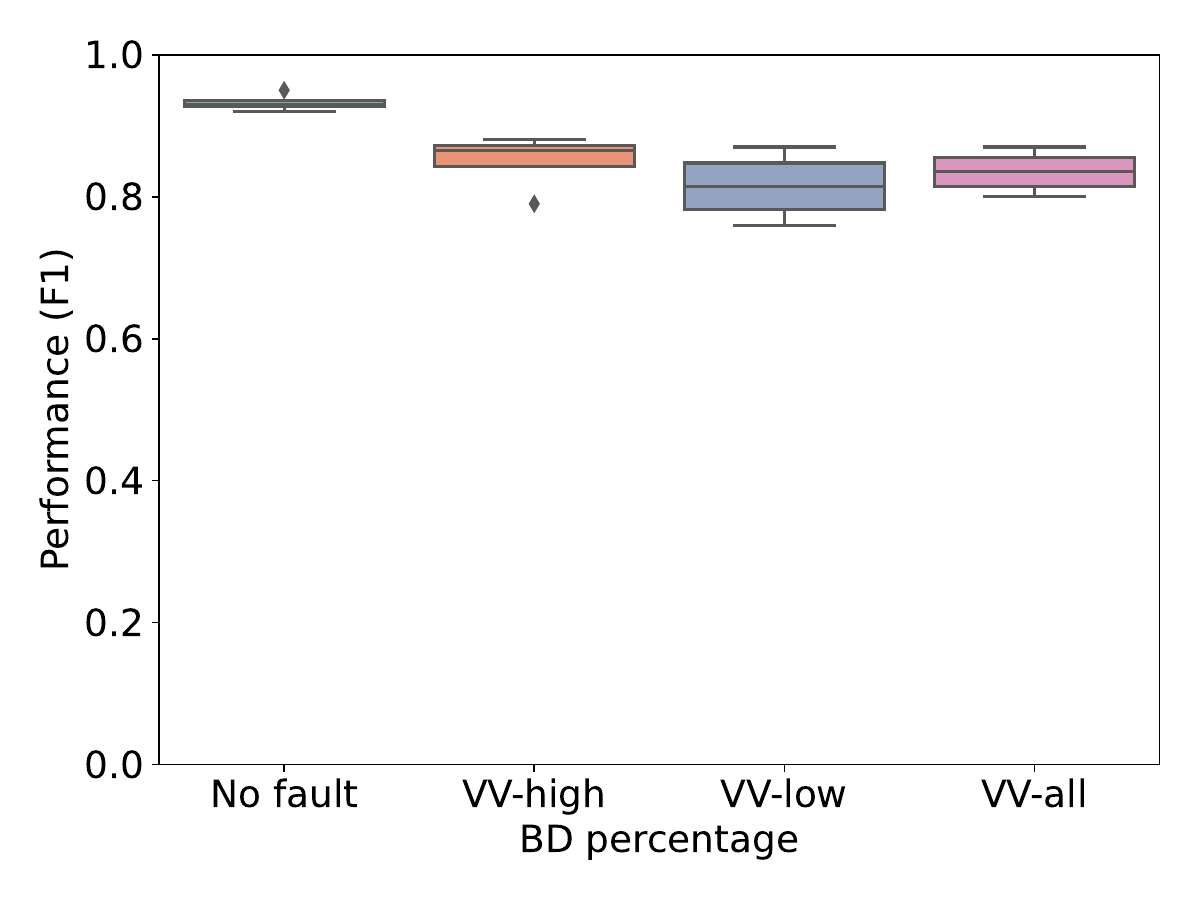}
    \caption{Performance of backdoor model on clean data}
    \label{fig:BackdoorPerf-1}
\end{figure}
\begin{figure}[t]
    \centering
    \includegraphics[width=0.9\linewidth]{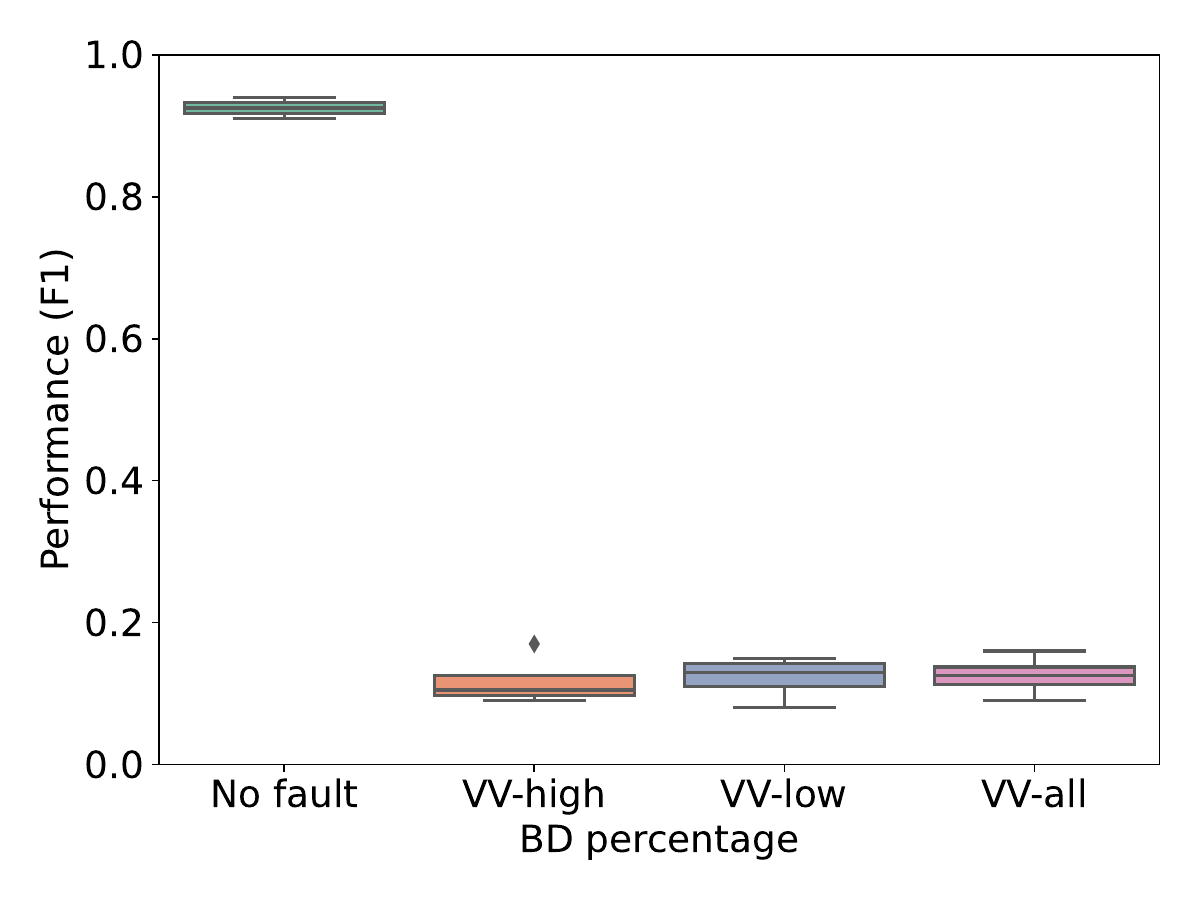}
    \caption{Performance of backdoor model on trigger data}
    \label{fig:BackdoorPerf-2}
\end{figure}

This section compares the performance of the backdoor model between clean and trigger data. For this, we conduct experiments to evaluate how the backdoored model performs when supplied with clean data and data with a trigger embedded individually. We apply the trigger to one class at a time for each experiment, run these experiments across four backdoor percentages (10, 20, 30, and 40\%), and compute the average and plot it.
As shown in Figure~\ref{fig:BackdoorPerf-1}, the model behaves normally when the clean data is sent to the model for classification, demonstrated by a high F1-score.
While in case of backdoor data, the model fails to classify as shown in in Figure~\ref{fig:BackdoorPerf-2} demonstrated by low F1-score.
This experiment proves that the attack is successful across all data classes and that the model is robust to the target class while preserving stealthiness.

In summary, the backdoor we design in this paper is realistic and flexible, and can be fine-tuned to control the utility-vs-stealthiness trade-off.


\section{Conclusion}

This paper proposes a backdoor attack targeting fault detection and localization in Cyber-Physical Systems(CPS). 
We first design a novel fault detection and localization model based on advanced LLM-based techniques, and we define a threat model, a trigger mechanism, and adversarial objectives for this model used in CPS. Our study shows that backdoor attacks pose severe threats because they remain dormant under normal conditions but enable targeted malicious behavior upon trigger activation. We design and evaluate a trigger generator and show that safeguarding the CPS-Artificial Intelligence (AI) integration pipeline and deployed infrastructure, and making them resilient against backdoor attacks, are essential for the safe deployment of AI-driven CPS.
The paper highlights the urgent need for secure and resilient AI-enabled fault management systems in future CPS deployments.
As a future extension of this work, we plan to evaluate various mitigation strategies, including robust training, trigger detection, runtime monitoring, and trusted learning architectures.  We also continue to evaluate the generalization properties of our technique across additional datasets and faults, which will enable the design of robust controls in the future.

\bibliographystyle{IEEEtran}
\bibliography{bib}

\end{document}